\documentclass[12pt]{article}

\usepackage{graphicx}
\usepackage{a4}

\def\U{\mathop{\rm U}}
\def\SU{\mathop{\rm SU}}
\def\SO{\mathop{\rm SO}}

\def\Spin{\mathop{\rm Spin}}

\newcommand{\wt}{\widetilde}

\usepackage{amsmath}
\usepackage{amsfonts}

\newcommand{\RR}{\mathbb{R}}
\newcommand{\CC}{\mathbb{C}}
\newcommand{\ZZ}{\mathbb{Z}}

\newcommand{\mm}{\mathfrak{m}}
\newcommand{\MM}{\mathfrak{M}}

\begin{document}

\begin{titlepage}
\title{\hfill\parbox{4cm}
       {\normalsize UT-09-05\\February 2009}\\
       \vspace{2cm}
Monopole operators in ${\cal N}=4$ Chern-Simons theories
and wrapped M2-branes
       \vspace{2cm}}
\author{
Yosuke Imamura\thanks{E-mail: \tt imamura@hep-th.phys.s.u-tokyo.ac.jp}
\\[30pt]
{\it Department of Physics, University of Tokyo,}\\
{\it Hongo 7-3-1, Bunkyo-ku, Tokyo 113-0033, Japan}
}
\date{}

\maketitle
\thispagestyle{empty}

\vspace{0cm}

\begin{abstract}
\normalsize
Monopole operators
in Abelian ${\cal N}=4$ Chern-Simons theories
described by circular quiver diagrams are investigated.
The magnetic charges of non-diagonal $\U(1)$ gauge symmetries
form the $\SU(p)\times\SU(q)$
root lattice where $p$ and $q$ are
numbers of untwisted and twisted hypermultiplets, respectively.
For monopole operators corresponding to the root vectors,
we propose a correspondence between the monopole operators and
states of a wrapped M2-brane in the dual geometry.
\end{abstract}

\end{titlepage}

\section{Introduction}
Since the proposal of Bagger-Lambert-Gusstavson (BLG) model\cite{Bagger:2006sk,Bagger:2007jr,Bagger:2007vi,Gustavsson:2007vu,Gustavsson:2008dy},
three-dimensional supersymmetric Chern-Simons theories
have attracted a great interest
as low energy effective theories of multiple M2-branes in
various backgrounds.
BLG model is $\SU(2)\times\SU(2)$
Chern-Simons theory
with bi-fundamental matter fields
which possesses ${\cal N}=8$ supersymmetry.
This is the first example of interacting Chern-Simons theory
with ${\cal N}\geq4$ supersymmetry.
Following the BLG model,
various Chern-Simons theories with ${\cal N}\geq4$
have been constructed
\cite{Gaiotto:2008sd,Fuji:2008yj,Hosomichi:2008jd,Aharony:2008ug,Hosomichi:2008jb,Bagger:2008se,Schnabl:2008wj,Imamura:2008dt},
and their properties have been studied intensively.

In this paper,
we discuss a field-operator correspondence in AdS$_4$/CFT$_3$.
In general field-operator correspondence claims that
there is one-to-one correspondence
between gauge invariant operators in CFT and fields in the
AdS space, and is 
one of most important claims of AdS/CFT correspondence.
For the ${\cal N}=6$ Chern-Simons theory,
Aharony-Bergman-Jafferis-Maldacena (ABJM) model\cite{Aharony:2008ug},
we need to take account of monopole operators
to reproduce the desired moduli spaces\cite{Aharony:2008ug}.
Namely, some of Kaluza-Klein modes on the gravity side
correspond to local operators carrying magnetic charges.
(See also \cite{Lambert:2008et,Distler:2008mk}
for similar analysis for BLG mode.)
Monopole operators in ABJM model are
further investigated in \cite{Berenstein:2008dc,Hosomichi:2008ip,Klebanov:2008vq}.

This is also the case in theories with less supersymmetries.
In the case of ${\cal N}=2$ quiver gauge theories
which describe M2-branes in toric Calabi-Yau $4$-folds,
the relation between
mesonic operators
and holomorphic monomial functions, which are specified by
the charges of toric $\U(1)$ symmetries,
was proposed in \cite{Lee:2007kv}.
In the reference, a simple prescription to
establish concrete correspondence between Kaluza-Klein modes
and mesonic operators is given by utilizing
brane crystals\cite{Lee:2006hw,Lee:2007kv,Kim:2007ic}.
When this method was proposed,
it had not been realized that
the quiver gauge theories are actually quiver Chern-Simons theories.
After the importance of the existence of Chern-Simons terms
was realized,
this proposal was confirmed\cite{Ueda:2008hx,Imamura:2008qs,Hanany:2008cd}
for special kind of brane crystals
which can be regarded as ``M-theory lift''
of brane tilings\cite{Hanany:2005ve,Franco:2005rj,Franco:2005sm}.
Monopole operators enter the correspondence again
as well as the case of ABJM model.
The results in \cite{Ueda:2008hx,Imamura:2008qs,Hanany:2008cd}
indicate, however,
that the set of primary operators corresponding to the
supergravity Kaluza-Klein modes
includes
only a special kind of monopole
operators, ``diagonal'' monopole operators,
which carries only the
diagonal $\U(1)$ magnetic charge and are constructed by combining
the dual photon fields and chiral matter fields.
(We now consider Abelian quiver Chern-Simons theories
and assume that the gauge group for each vertex is $\U(1)$.)

The other monopole operators, which we call non-diagonal monopole operators,
have no correspondents in the bulk Kaluza-Klein modes.
In \cite{Imamura:2008ji}, it is suggested that such non-diagonal monopole operators
correspond to M2-branes wrapped on $2$-cycles in the internal space.
The purpose of this paper is to study this correspondence in more detail
for ${\cal N}=4$ Abelian quiver Chern-Simons theories
described by
circular quiver diagrams\cite{Hosomichi:2008jd,Imamura:2008dt}.

Because we consider Abelian Chern-Simons theory,
whose gauge group is the product of $\U(1)$,
the dual geometry has large curvature.
By this reason, we mainly focus only on
the charges of global symmetries,
which are quantized and are hopefully
reproduced on the gravity side correctly.
We do not attempt to reproduce the conformal dimension
of monopole operators by using the gravity description.

This paper is organized as follows.
In the next section we briefly explain the relation between
the dual photon field and monopole operators
in quiver Chern-Simons theories.
In \S\ref{radial.sec} we review the radial quantization
method used in \cite{Borokhov:2002ib,Borokhov:2002cg}
to compute the conformal dimension and global $\U(1)$ charges
of monopole operators.
After explaining the ${\cal N}=4$ Chern-Simons
theory in \S\ref{n4.sec}
and the structure of the dual geometry in \S\ref{int.sec},
we discuss the duality between non-diagonal primary monopole operators
and wrapped M2-branes in \S\ref{pmo.sec}.
The last section is devoted to conclusions and discussions.

\section{Monopole operators and the dual photon field}\label{ndp.sec}
To briefly review basic facts about monopole operators,
let us consider a generic ${\cal N}=2$ quiver Chern-Simons theory
described by a connected quiver diagram with $n$ vertices.
We label vertices and edges by $a$ and $I$, respectively.
We assume that the gauge group of every vertex is $\U(1)$.
We denote the gauge group for vertex $a$ by $\U(1)_a$,
and its Chern-Simons level by $k_a$.
We impose the constraint
\begin{equation}
\sum_{a=1}^nk_a=0,
\label{ktot}
\end{equation}
on the levels
to obtain moduli space which can be regarded as the background
of an M2-brane.
The action includes the following Chern-Simons terms.
\begin{equation}
S_{\rm CS}=\sum_{a=1}^n\frac{k_a}{4\pi}A_adA_a.
\label{uncs}
\end{equation}
We define another basis for $n$ $\U(1)$ gauge fields.
We recombine $A_a$ into $n$ gauge fields
\begin{equation}
A_D, A_B, A'_1,\ldots,A'_{n-2}.
\label{dbbasis}
\end{equation}
$A_D$ is the gauge field of $\U(1)_D$,
the diagonal $\U(1)$ subgroup.
When we represent $A_a$ as linear combinations
of gauge fields in (\ref{dbbasis}),
$A_D$ enters all of them with coefficient $1$:
\begin{equation}
A_a=A_D+\cdots,
\end{equation}
where $\cdots$ represents linear combinations of
$A_B$ and $A_i'$.
By substituting this into
(\ref{uncs}), we obtain
\begin{equation}
S_{\rm CS}=\sum_{a=1}^n\frac{1}{2\pi}A_DdA_B+\cdots,
\label{scsadb}
\end{equation}
where $\cdots$ does not includes $A_D$.
Thanks to (\ref{ktot}) we do not have $A_DdA_D$ term.
$A_B$ in (\ref{dbbasis}) is defined by this equation
as the gauge field appearing in the
linear term of $A_D$,
and is given by
\begin{equation}
A_B=\sum_{a=1}^nk_aA_a.
\label{defab}
\end{equation}
The diagonal gauge field $A_D$ does not couple to
matter fields and appears only in the Chern-Simons term
(\ref{scsadb}), and
the equation of motion of $A_D$ is
\begin{equation}
dA_B=0.
\label{dab0}
\end{equation}
Due to the ``pure gauge'' constraint (\ref{dab0}),
we can define the dual photon field $a$ by
\begin{equation}
A_B=da.
\label{dpdef}
\end{equation}

The dual photon field is periodic field
with the period $2\pi$\cite{Martelli:2008si},
and
it is convenient to define operators in the form
\begin{equation}
e^{ima},\quad m\in\ZZ.
\label{eima}
\end{equation}
Because the $\U(1)_D$ field strength $F_D$ is the canonical conjugate of
the operator $a$,
the operator (\ref{eima}) shifts the $\U(1)_D$ flux by $m$.
In other words,
this operator carries the magnetic charge $m$ for every $\U(1)_a$.
We call such operators
diagonal monopole operators.
General diagonal monopole operators can be constructed by
combining $e^{ima}$ and other magnetically neutral operators.

The relations (\ref{defab}) and (\ref{dpdef}) indicate that
the dual photon field $a$ is transformed
under a gauge transformation $\delta A_a=d\lambda_a$
by
\begin{equation}
\delta a=\sum_{a=1}^nk_a\lambda_a.
\end{equation}
This means that the operator $e^{ima}$ carries
electric $\U(1)_a$ charge $mk_a$.

There also exist monopole operators which carry
non-diagonal magnetic charges.
Let $m_a$ be the $\U(1)_a$ magnetic charge of
an operator.
The equation of motion of $A_a$ is
\begin{equation}
k_aF_a+j_a=0,
\end{equation}
where $j_a$ is the matter contribution to the electric $\U(1)_a$ current.
By integrating this equation over a sphere enclosing
the operator, we obtain
\begin{equation}
k_am_a+Q_a=0,
\label{gauss}
\end{equation}
where $Q_a$ is the matter contribution to the $\U(1)_a$ charge
of the operator.
This is the Gauss law constraint guaranteeing the gauge invariance of the
operator.
(Because we consider only rotationally invariant operators,
this integrated form is sufficient to guarantee their gauge invariance.)

The magnetic charges are constrained by
the equation
\begin{equation}
\sum_{a=1}^nk_am_a=0,
\label{macponsyt}
\end{equation}
which is obtained by integrating (\ref{dab0})
or summing up (\ref{gauss}) over $a$.
Because of this constraint
the number of independent non-diagonal monopole charges is $n-2$.
In the case of ${\cal N}=4$ theory,
this number is indeed the same as
two-cycles in the internal space\cite{Imamura:2008ji}.

\section{Radial quantization method}\label{radial.sec}
We use the radial quantization method\cite{Borokhov:2002ib,Borokhov:2002cg}
to compute the conformal dimension
and global $\U(1)$ charges of monopole operators.
We want to look for operators saturating the BPS bound
\begin{equation}
\Delta\geq R,
\label{bpsbound}
\end{equation}
where $R$ is the charge of $\U(1)_R$ subgroup of the ${\cal N}=2$ superconformal
group.

We map an Euclidean three-dimensional CFT in $\RR^3$
to the theory in ${\bf S}^2\times\RR$ by a conformal transformation.
A monopole operator with magnetic charges $m_a$ corresponds to a state
in the Hilbert space defined in ${\bf S}^2$ with flux $m_a$
through it.
The conformal dimension of the operator is computed as
the energy of the corresponding state.
We can also obtain $\U(1)$ charges of monopole operators
as the charges of the corresponding states.

The fields in vector multiplets are treated as classical background fields.
We expand fields in the chiral multiplets into spherical harmonics,
and define creation and annihilation operators,
which are used to construct the Hilbert space.
Mode expansion of scalar and spinor fields
in BPS monopole backgrounds is given in \cite{Borokhov:2002cg}.
Let $\mu\in\ZZ$ be the number of the flux
coupling to a chiral multiplet $\Phi=(\phi,\psi)$.
The scalar component $\phi$ and the fermion component
$\psi$
are expanded by
\begin{align}
\phi
&=\sum_{l=\frac{|\mu|}{2}}^\infty\sum_{m=-l}^l \alpha_{l,m} e^{-(l+1/2)\tau}Y^0_{l,m}
+\sum_{l=\frac{|\mu|}{2}}^\infty\sum_{m=-l}^l \beta^\dagger_{l,m} e^{(l+1/2)\tau}Y^0_{l,m},\label{modeexp1}
\\
\psi
&=\sum_{l=\frac{|\mu|+1}{2}}^\infty\sum_{m=-l}^l a_{l,m} e^{-(l+1/2)\tau}Y^+_{l,m}
+\sum_{l=\frac{|\mu|-1}{2}}^\infty\sum_{m=-l}^l b^\dagger_{l,m} e^{(l+1/2)\tau}Y^-_{l,m},
\label{modeexp2}
\end{align}
where $Y^0_{l,m}$ and $Y^\pm_{l,m}$ are spherical harmonics
of scalar and spinor on the ${\bf S}^2$ with flux.
Refer to \cite{Borokhov:2002cg} for more detail.
To obtain the expansion above, we used the free field equations.
The radial quantization method with the expansion
(\ref{modeexp1}) and (\ref{modeexp2}) gives the tree level conformal dimensions
for $\phi$ and $\psi$, and this cannot be justified in
general ${\cal N}=2$ theories in which the conformal dimension recieves
large quantum corrections.
In the ${\cal N}=4$ theory the conformal dimensions of primary operators
are protected by the non-abelian R-symmetry, and we assume that
the applicability of the free field approximation in the
computation below.

All the oscillators $\alpha_{l,m}$, $\beta_{l,m}$, $a_{l,m}$, and $b_{l,m}$
have the same indices $l$ (angular momentum) and $m$ (magnetic quantum number)
associated with the rotational symmetry of ${\bf S}^2$.
$l$ must be non-negative, and
when $\mu=0$ the term including $b^\dagger_{-1/2,m}$ should be omitted.
The energy of a quantum for each oscillator is
\begin{equation}
E_l=l+\frac{1}{2},
\end{equation}
for any of four kinds of oscillators.

The conformal dimension of the monopole operator corresponding to the Fock vacuum
is computed as the zero-point energy.
By using an appropriate regularization,
we obtain the contribution of the oscillators
of $\phi$ and $\psi$ as
\begin{equation}
\Delta_0=\frac{|\mu|}{4}.
\label{eforphi}
\end{equation}

We can also compute $\U(1)$ charges of the monopole operator.
If a $\U(1)$ charge of the fermion $\psi$ in a chiral multiplet is $q$,
the contribution of the chiral multiplet to the zero point charge is
\begin{equation}
Q_0=-\frac{|\mu|q}{2}.\label{qforphi}
\end{equation}

Excited states are constructed by acting creation operators
on the Fock vacuum.
If we assume that the R-charge
of chiral multiplets is not corrected from
the classical value,
only creation operator saturating
the BPS bound (\ref{bpsbound}) is $\beta^\dagger_{0,0}$,
and it exists only when $\mu=0$.
We can use only this operator to construct excited BPS states.

$\Delta_0$ and $Q_0$ in a quiver gauge theory is
obtained by summing
up the contribution of all chiral multiplets.
Let $Q_{aI}$ be the $\U(1)_a$ charge of
chiral multiplet $\Phi_I$.
We consider a monopole operator
with magnetic $\U(1)_a$ charge $m_a$.
The flux coupling to $\Phi_I$ is
given by
\begin{equation}
\mu_I=\sum_{a=1}^n m_aQ_{aI}.
\end{equation}
The energy of the Fock vacuum is
\begin{equation}
\Delta_0=\frac{1}{4}\sum_I|\mu_I|.
\end{equation}
The summation is taken over all the bi-fundamental
chiral multiplets.
For a $\U(1)$ symmetry,
if the charge of chiral multiplet $\Phi_I$ is $q_I$,
the zero-point charge of the $\U(1)$ symmetry is
\begin{equation}
Q_0=-\frac{1}{2}\sum_I|\mu_I|q_I
\end{equation}

For the R-symmetry, $q_I=-1/2$, and
$Q_0$ coincides with $\Delta_0$.
Namely, the BPS bound (\ref{bpsbound}) is saturated by the vacuum state.
General BPS states are constructed by
acting the creation operators
$\beta^\dagger_{I,0,0}$, which exist only for chiral fields
with $\mu_I=0$, on the Fock vacuum.

\section{${\cal N}=4$ Chern-Simons theory}\label{n4.sec}
\begin{figure}[htb]
\centerline{\includegraphics{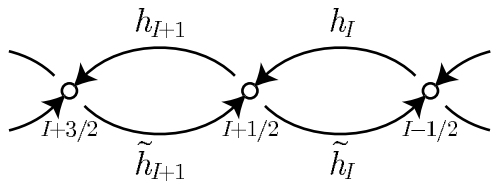}}
\caption{A part of a circular quiver diagram of an ${\cal N}=4$ supersymmetric Chern-Simons theory is shown.
Arrows represent chiral multiplets.}
\label{quiver.eps}
\end{figure}

Let us consider an Abelian ${\cal N}=4$ Chern-Simons
theory described by a circular quiver diagram\cite{Hosomichi:2008jd,Imamura:2008dt}
with period $n$ shown in Figure \ref{quiver.eps}.

We label hypermultiplets by integers $I$ in order in the quiver diagram.
$I$ is defined only modulo $n$ and $I=1$ and $I=n+1$ are identified.
In terms of the language of ${\cal N}=2$ supersymmetry,
a hypermultiplet $I$ consists of two chiral multiplets,
$h_I$ and $\wt h_I$.
We use half odd integers to label vertices,
and denote $\U(1)$ gauge symmetry coupling to $h_I$ and $h_{I+1}$
by $\U(1)_{I+\frac{1}{2}}$.
$\U(1)_{I+\frac{1}{2}}\times\U(1)_{I-\frac{1}{2}}$ charges of
$h_I$ and $\wt h_I$ are $(+1,-1)$ and $(-1,+1)$, respectively.

There are two
kinds of hypermultiplets,
which are called untwisted and twisted hypermultiplets \cite{Hosomichi:2008jd}.
Let us define numbers $s_I$ associated with hypermultiplets
which are $0$ for untwisted hypermultiplets and $1$ for twisted hypermultiplets.
\begin{equation}
s_I=0: \mbox{untwisted hypermultiplet},\quad
s_I=1: \mbox{twisted hypermultiplet}.
\end{equation}
We use indices $a, b,\ldots$ to label untwisted hypermultiplets and
$\dot a, \dot b,\ldots$ for twisted hypermultiplets.
Namely, $a$ ($\dot a$) runs over integers $I$ such that $s_I=0$ ($s_I=1$).

This theory possesses
the R-symmetry
\begin{equation}
\Spin(4)_R=\SU(2)_A\times\SU(2)_B,
\end{equation}
and flavor symmetry
\begin{equation}
\U(1)_A\times\U(1)_B.
\end{equation}
We denote the generators of $\SU(2)_A$, $\SU(2)_B$, $\U(1)_A$, and $\U(1)_N$
by $T_i$, $\wt T_i$ ($i=1,2,3$), $P$, and $\wt P$, respectively.

Scalar fields in untwisted and twisted hypermultiplets
are transformed by $\SU(2)_A$ and $\SU(2)_B$, respectively.
We can form the doublets as
\begin{equation}
h_a^\alpha
=\left(\begin{array}{c}
h_a \\ \wt h_a^*
\end{array}\right),\quad
h_{\dot a}^{\dot\alpha}
=\left(\begin{array}{c}
h_{\dot a} \\ \wt h_{\dot a}^*
\end{array}\right),
\end{equation}
where $\alpha$ and $\dot\alpha$ are $\SU(2)_A$ and $\SU(2)_B$ spinor indices,
respectively.
The conformal dimension $\Delta$ and the charges $T_3$, $P$, $\wt T_3$, and $\wt P$ of
scalar fields are shown in
Table \ref{myper.tbl}.
\begin{table}[htb]
\caption{The conformal dimension and charges of scalar components of multiplets are shown.}
\label{myper.tbl}
\begin{center}
\begin{tabular}{cccccc}
\hline
\hline
    & $\Delta$ & $T_3$ & $P$ & $\wt T_3$ & $\wt P$ \\
\hline
$h_a$ & $\frac{1}{2}$ & $\frac{1}{2}$ & $1$ & $0$ & $0$ \\
$\wt h_a$ & $\frac{1}{2}$ & $\frac{1}{2}$ & $-1$ & $0$ & $0$ \\
$h_{\dot a}$ & $\frac{1}{2}$ & $0$ & $0$ & $\frac{1}{2}$ & $1$ \\
$\wt h_{\dot a}$ & $\frac{1}{2}$ & $0$ & $0$ & $\frac{1}{2}$ & $-1$ \\
\hline
\end{tabular}
\end{center}
\end{table}
The R-charge of the ${\cal N}=2$ superconformal subgroup is
\begin{equation}
R=T_3+\wt T_3,
\end{equation}
and all the scalar components of the chiral multiplets
saturate the BPS bound
\begin{equation}
\Delta\geq T_3+\wt T_3.
\label{rt3t3}
\end{equation}

In order for the theory to possess ${\cal N}=4$ supersymmetry,
the levels should be given by
\begin{equation}
k_{I+\frac{1}{2}}=k(s_{I+1}-s_I),\quad
k\in\ZZ.
\label{kiis}
\end{equation}
We refer to the integer $k$ simply as the ``level'' of the theory.

The Higgs branch moduli space
of this theory is analyzed in \cite{Imamura:2008nn}.
See also \cite{Benna:2008zy,Terashima:2008ba}.
When $k=1$, it is the product of two orbifolds
\begin{equation}
{\cal M}_{p,q}
=\CC^2/\ZZ_p
\times\CC^2/\ZZ_q,
\label{orbi}
\end{equation}
where $p$ and $q$ are the numbers of untwisted and twisted
hypermultiplets, respectively.
The complex coordinates of the $\CC^2/\ZZ_p$ factor
can be spanned by
\begin{align}
M=h_a\wt h_a,\quad
X=e^{-ia}\prod_ah_a,\quad
\wt X=e^{ia}\prod_a\wt h_a.
\label{mxy}
\end{align}
The operator $M$ is independent of the index $a$ due to the F-term conditions.
By definition, these three operators satisfy $M^p=X\wt X$, and
this is nothing but the defining equation of the orbifold $\CC^2/\ZZ_p$.
The generator of the orbifold group $\ZZ_p$
which keeps
the operators in (\ref{mxy})
invariant is
\begin{equation}
e^{2\pi iP/p}\in\U(1)_A.
\label{Zp}
\end{equation}

The other factor $\CC^2/\ZZ_q$ in (\ref{orbi}) is parameterized by
\begin{equation}
N=h_a\wt h_a,\quad
Y=e^{ia}\prod_{\dot a}h_{\dot a},\quad
\wt Y=e^{-ia}\prod_{\dot a}\wt h_{\dot a},
\label{mxy2}
\end{equation}
and these satisfy $N^q=Y\wt Y$, the defining equation of $\CC^2/\ZZ_q$.
The generator of $\ZZ_q$ is
\begin{equation}
e^{-2\pi i\wt P/q}\in\U(1)_B.
\label{Zq}
\end{equation}

When $k\geq2$, the electric charges of the operator $e^{ia}$
becomes $k$ times those for $k=1$.
In this case, we formally define $(M,X,Y)$ and $(\wt M,\wt X,\wt Y)$ by
(\ref{mxy}) and (\ref{mxy2}) with $e^{\pm ia}$ replaced by $e^{\pm ia/k}$.
Because $e^{\pm ia/k}$ is ill defined due to the fractional coefficient
in the exponent,
we need to combine these formal operators
so that the coefficient in the exponent becomes integral.
This is equivalent to imposing the invariance under
\begin{equation}
(X,Y,\wt X,\wt Y)
\rightarrow
(\omega_kX,\omega_k^{-1}Y,\omega_k^{-1}\wt X,\omega_k\wt Y).
\end{equation}
This transformation is realized by
\begin{equation}
e^{2\pi i(P/kp-\wt P/kq)}\in
\U(1)_A\times\U(1)_B.
\label{Zk}
\end{equation}
This means that the
moduli space is orbifold of (\ref{orbi})
divided by $\ZZ_k$ generated by (\ref{Zk}).
As the result we obtain the orbifold
\begin{equation}
{\cal M}_{p,q,k}
=((\CC^2/\ZZ_p)\times(\CC^2/\ZZ_q))/\ZZ_k.
\end{equation}

\section{Internal space}\label{int.sec}
The gravity dual of the ${\cal N}=4$ Chern-Simons theory
is $AdS_4\times X_7$ with
\begin{equation}
X_7
={\cal M}_{k,p,q}|_{r=1}
=({\bf S}^7/(\ZZ_p\times\ZZ_q))/\ZZ_k.
\end{equation}
The homologies $H_i(X_7,\ZZ)$ of this manifold are\cite{Imamura:2008ji}
\begin{eqnarray}
&&
H_0=\ZZ,\quad
H_1=\ZZ_k,\quad
H_2=\ZZ^{p+q-2},\quad
H_3=(\ZZ_{kp}^{q-1}\times\ZZ_{kq}^{p-1}\times\ZZ_{kpq})/(\ZZ_{p}\times\ZZ_q),
\nonumber\\&&
H_4=0,\quad
H_5=\ZZ^{p+q-2}\times\ZZ_k,\quad
H_6=0,\quad
H_7=\ZZ.
\label{homology2}
\end{eqnarray}
In order to discuss the relation between monopole
operators and wrapped M2-branes in $X_7$,
we need to know where two- and three-cycles are
in the manifold $X_7$.
For this purpose it is convenient to
represent $X_7$ as a ${\bf T}^2$ fibration over $B={\bf S}^5$
by using the global symmetry $\U(1)_A\times\U(1)_B$ to
define ${\bf T}^2$ fibers as orbits.

\begin{figure}[htb]
\centerline{\includegraphics{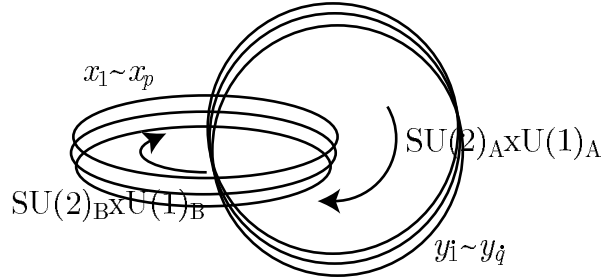}}
\caption{The loci $x_a$ and $y_{\dot a}$ in the base manifold ${\bf S}^5$
are shown.
On the loci $x_a$ $\SU(2)_A\times\U(1)_A$ acts as isometry while
$\SU(2)_B\times\U(1)_B$ does as transverse rotations.
For $y_{\dot a}$ the roles of these symmetries are exchanged.}
\label{3cycles.eps}
\end{figure}

Let us first describe the covering space $\wt X_7={\bf S}^7$ as
${\bf T}^2$ fibration over $B$.
Each of $\U(1)_A$ and $\U(1)_B$ has fixed submanifold ${\bf S}^3\subset{\bf S}^7$.
We denote those for $\U(1)_A$ and $\U(1)_B$ by ${\bf S}^3_A$ and ${\bf S}^3_B$,
respectively.
${\bf S}^3_A$ and ${\bf S}^3_B$ are projected into two ${\bf S}^2\subset B$,
${\bf S}^2_A$ and ${\bf S}^2_B$, linking with each other.
By the $\ZZ_p\subset\U(1)_A$ orbifolding and blowing up the resultant
orbifold singularity,
${\bf S}^2_A$ split into $p$ loci in $B$, which we refer to as
$x_a$ ($a=1,\ldots,p$).
Similarly, the $\ZZ_q\subset\U(1)_B$ orbifolding and the blow-up generate $q$ loci,
$y_{\dot a}$ ($\dot a=\dot1,\ldots,\dot q$).
(Figure \ref{3cycles.eps})
(Although we blew up the singularities to define
the loci $x_a$ and $y_{\dot a}$, we only consider the singular limit.)
We use indices $a$ and $\dot a$ for the loci
just like the two types of hypermultiplets.
As is mentioned in \cite{Imamura:2008ji},
by a certain duality between M2-branes in the orbifold
and a D3-fivebrane system in type IIB string theory,
the loci are mapped to
fivebranes,
and each hypermultiplet arises at the intersection of each fivebrane
and D3-branes.
Through this duality, we have a natural one-to-one correspondence
between the loci and the hypermultiplets.

We define
$\alpha$-, $\beta$-, and $\gamma$-cycles in the ${\bf T}^2$ fiber,
as cycles corresponding to the generators
$e^{2\pi iP/p}$ in (\ref{Zp}),
$e^{-2\pi i\wt P/q}$ in (\ref{Zq}),
and
$e^{2\pi i(P/kp-\wt P/kq)}$ in (\ref{Zk}),
respectively.
The $\alpha$- ($\beta$-)cycle shrinks on the loci $x_a$ ($y_{\dot a}$).
The two-cycle homology $H_2(X_7,\ZZ)$ is generated by
\begin{equation}
[x_a,x_b]^\alpha,\quad
[y_{\dot a},y_{\dot b}]^\beta,
\end{equation}
where $[x_a,x_b]$ represents a segment in $B$ connecting two loci,
$x_a$ and $x_b$,
and the superscript $\alpha$ means the lift of the segment to
the two-cycle in $X_7$ by combining the $\alpha$-cycle.
$[y_{\dot a},y_{\dot b}]^\beta$ is defined similarly.
It is convenient to define
the formal basis ${\bf x}_a$ and ${\bf y}_{\dot a}$
by $[x_a,x_b]={\bf x}_a-{\bf x}_b$ and so on.
The general two-cycles are in the form
\begin{equation}
\Sigma_2
=\sum_ac_a{\bf x}_a^\alpha
+\sum_{\dot a}c_{\dot a}{\bf y}_{\dot a}^\beta,\quad
c_a,c_{\dot a}\in\ZZ,
\label{twocycles}
\end{equation}
with the coefficients satisfying
\begin{equation}
\sum_ac_a=\sum_{\dot a}c_{\dot a}=0.
\label{cconst}
\end{equation}

A set of generating $3$-cycles of $H_3(X_7,\ZZ)$ is
\begin{equation}
[x_a,x_b]^{\alpha\gamma}
={\bf x}_a^{\alpha\gamma}-{\bf x}_b^{\alpha\gamma},\quad
[y_{\dot a},y_{\dot b}]^{\alpha\gamma}
={\bf y}_{\dot a}^{\alpha\gamma}-{\bf y}_{\dot b}^{\alpha\gamma},\quad
[x_a,y_{\dot b}]^{\alpha\gamma}
={\bf x}_a^{\alpha\gamma}-{\bf y}_{\dot b}^{\alpha\gamma}.
\label{threekin2}
\end{equation}
The superscripts ``$\alpha\gamma$'' mean the lift of segments
to $3$-cycles by combining $\alpha$- and $\gamma$-cycles with the segments.
The $3$-cycle homology group is defined as the set of
elements in the form
\begin{equation}
\Sigma_3=\sum n_a{\bf x}_a^{\alpha\gamma}+\sum n_{\dot a}{\bf y}_{\dot a}^{\alpha\gamma},\quad
n_a,n_{\dot a}\in\ZZ,
\end{equation}
with the coefficients constrained by
\begin{equation}
\sum_an_a
+\sum_{\dot a}n_{\dot a}
=0,
\end{equation}
and the identification relations
\begin{equation}
k{\bf v}_a^{\alpha\gamma}
=k{\bf w}_{\dot a}^{\alpha\gamma}
=0,
\end{equation}
where ${\bf v}_a$ and ${\bf w}_{\dot a}$ are defined by
\begin{equation}
{\bf v}_a=-q{\bf x}_a+\sum_{\dot b=\dot1}^{\dot q}{\bf y}_{\dot b},\quad
{\bf w}_{\dot a}=\sum_{b=1}^p{\bf x}_b-p{\bf y}_{\dot a}.
\label{vwdef}
\end{equation}

\section{Monopole operators and M2-branes}\label{pmo.sec}
Monopole operators are
labeled by $n$ magnetic charges $m_{I+\frac{1}{2}}\in\ZZ$.
We define the group of non-diagonal magnetic charges
as the set of charges $m_{I+\frac{1}{2}}$ with identification
\begin{equation}
(m_{\frac{1}{2}},\cdots,m_{n-\frac{1}{2}})
\sim(m_{\frac{1}{2}}+1,\cdots,m_{n-\frac{1}{2}}+1)
\end{equation}
removing the diagonal $\U(1)$ charge.
In order to realize this identification automatically,
we use the relative charges $\mu_I$ defined by
\begin{equation}
\mu_I=m_{I+\frac{1}{2}}-m_{I-\frac{1}{2}}.
\end{equation}
This can be regarded as the effective flux for
hypermultiplet $I$.
By definition,
$\mu_I$ are constrained
by
\begin{equation}
\sum_I\mu_I=0.
\label{musum}
\end{equation}
(\ref{macponsyt})
imposes further constraint
\begin{equation}
0=\sum_Ik_{I+\frac{1}{2}}m_{I+\frac{1}{2}}
=-k\sum_Is_I\mu_I.
\label{n4km}
\end{equation}
(\ref{musum}) and (\ref{n4km}) can be rewritten in
the following form.
\begin{equation}
\sum_a\mu_a=\sum_{\dot a}\mu_{\dot a}=0.
\label{muconst}
\end{equation}
The integers $\mu_I$ satisfying (\ref{muconst}) form
the $\SU(p)\times\SU(q)$ root lattice.

There are $n-2$ independent charges and we would like to identify these
with the wrapping charges of M2-branes.
Indeed, the two-cycle Betti number of the internal space $X_7$
is $b_2=n-2$ and coincides
the number of non-diagonal magnetic charges.
Not only the coincidence of the numbers of charges,
we want to establish the one to one correspondence between
the magnetic charges $\mu_I$
and two-cycles in (\ref{twocycles}).
A natural guess consistent with (\ref{cconst}) is
\begin{equation}
\Sigma_2[\mu_I]
=\sum_a\mu_a{\bf x}_a^\alpha+\sum_{\dot a}\mu_{\dot a} {\bf y}_{\dot a}^\beta.
\label{sigma2i}
\end{equation}

Let us consider magnetic operators which are primary in the sense of ${\cal N}=2$
superconformal symmetry.
This means that we look for operators saturating (\ref{rt3t3}).

The zero-point contribution to the conformal dimension and the R-charge are
\begin{equation}
\Delta_0=R_0=\frac{1}{2}\sum_I|\mu_I|.
\end{equation}
For simplicity, we consider operators with minimum values of $R_0$.
Because $\mu_I$ is constrained by (\ref{muconst}),
the minimum $R_0$ is $1$ for monopoles
with one relative charge $+1$ and one relative charge $-1$.
The indices of the two non-vanishing relative charges
should be both undotted or both dotted.
Namely, there are the following two sets of monopole operators
\begin{align}
\mm_{ab}:\quad
\mu_c=-\delta_{ca}+\delta_{cb},\quad
\mu_{\dot c}=0,\\
\mm_{\dot a\dot b}:\quad
\mu_c=0,\quad
\mu_{\dot c}=-\delta_{\dot c\dot a}+\delta_{\dot c\dot b}.
\end{align}
The conformal dimensions and global $\U(1)$ charges
of these operators are given in Table \ref{mmcharges.tbl}.
\begin{table}[htb]
\caption{The conformal dimension and global $\U(1)$ charges of
monopole operators $\mm_{ab}$ and $\mm_{\dot a\dot b}$ are shown.}
\label{mmcharges.tbl}
\begin{center}
\begin{tabular}{cccccc}
\hline
\hline
    & $\Delta$ & $T_3$ & $P$ & $\wt T_3$ & $\wt P$ \\
\hline
$\mm_{ab}$ & $1$ &  $1$ & $0$ & $0$ & $0$ \\
$\mm_{\dot a\dot b}$ & $1$ & $0$ & $0$ & $1$ & $0$ \\
\hline
\end{tabular}
\end{center}
\end{table}

Because two sets are discussed in parallel way,
we focus only on the operators $\mm_{ab}$ in the following.

The magnetic charges of monopole operators $\mm_{ab}$ form
$\SU(p)$ root system.
Indeed, the intersection among the cycles (\ref{sigma2i}) for $\mm_{ab}$
forms the
$\SU(p)$ Cartan matrix.
In the dual geometry this $\SU(p)$ can be identified with the
gauge symmetry on the coincident $p$ D6-branes,
which arises from the $\CC^2/\ZZ_p$ singularity
through the $\U(1)_A$ orbit compactification to type IIA string theory.
If we identify the wrapped M2-branes with
the non-diagonal components of the $\SU(p)$ vector multiplets
on the D6-branes,
we can interpret the charge $T_3[\mm_{ab}]=1$ as the R-charge of a scalar field
in the vector multiplet.
Because $\SU(2)_A$
from the type IIA perspective is the transverse rotation
around the D6-branes,
the scalar components of the vector multiplet belong
to the $\SU(2)_A$ triplet.
There is one component with $T_3=1$, and is identified with the
operator $\mm_{ab}$.

In general, the vacuum state does not give the
gauge invariant operators.
For the operator to be gauge invariant,
the Gauss law constraint (\ref{gauss})
must be satisfied.
The (absolute) magnetic charges $m_{I+\frac{1}{2}}$ of the
monopole operator $\mm_{ab}$ is given by
\begin{equation}
m_{I+\frac{1}{2}}[\mm_{ab}]=d+\left[a>I+\frac{1}{2}>b\right]
\label{abscharge}
\end{equation}
where $d$ is an arbitrary integer representing the diagonal magnetic
charge,
and the inequality in the bracket stands for $1$ ($0$) if it is true (false).
Because the quiver diagram is circular,
we cannot say which of given
two indices, say $a$ and $b$, is greater or smaller.
However, we can say if three indices are in the descending order or not.
In this sense, the bracket in
(\ref{abscharge}) is well defined.

In order to satisfy (\ref{gauss})
we need to add an appropriate set of chiral multiplets.
Gauge invariant monopole operators are given by
\begin{equation}
\MM_{ab}
=\left\{
\begin{array}{l}
\displaystyle
\mm_{ab}{\cal O}_{\rm neutral}\prod_{a>\dot c>b}h_{\dot c}^{k(d+1)}
\prod_{b>\dot c>a}h_{\dot c}^{kd}\quad
(d\geq0), \\
\displaystyle
\mm_{ab}{\cal O}_{\rm neutral}\prod_{a>\dot c>b}\wt h_{\dot c}^{-k(d+1)}
\prod_{b>\dot c>a}\wt h_{\dot c}^{-kd}\quad
(d\leq-1),
\end{array}\right.
\end{equation}
where 
${\cal O}_{\rm neutral}$.
is an electrically and magnetically neutral operator.
The products are taken with respect to $\dot c$ satisfying the inequalities
in the sense we mentioned above.
Note that we cannot use $h_a$ and $h_b$ because
when $\mu_I\neq0$ the corresponding chiral multiplet
does not include oscillators saturating the BPS bound.
Due to the F-term conditions
${\cal O}_{\rm neutral}$ can be written in terms of
$M$ in (\ref{mxy}) and $N$ in (\ref{mxy2}) by
\begin{equation}
{\cal O}_{\rm neutral}=M^mN^n,\quad
m,n=0,1,2,\ldots.
\end{equation}
By using charges given in
Tables \ref{myper.tbl} and \ref{mmcharges.tbl},
we obtain the following charges of $\MM_{ab}$:
\begin{align}
T_3[\MM_{ab}]
=&1+m,& m=&0,1,2,\ldots,
\label{t3m}\\
P[\MM_{ab}]
=&0,\\
\wt T_3[\MM_{ab}]
=&\frac{1}{2}|\wt P[\MM_{ab}]|+n,& n=&0,1,2,\ldots,
\label{wtt3mm}\\
\wt P[\MM_{ab}]
=&\left(d+\frac{q_{[a,b]}}{q}\right)kq,& d=&0,\pm1,\pm2,\ldots,
\label{kkmomcon}
\end{align}
where $q_{[a,b]}$ is the number of twisted hypermultiplets
between
untwisted hypermultiplets $a$ and $b$ in the quiver diagram.
Namely,
by using the bracket used in (\ref{abscharge}),
$q_{[a,b]}$ is given by
\begin{equation}
q_{[a,b]}=\sum_{\dot c}[a>\dot c>b].
\end{equation}

Let us interpret these charges in terms of wrapped M2-branes
in the dual geometry.
Wrapped M2-branes are localized on the $\U(1)_A$ fixed submanifold.
It is the Lens space $L_{kq}={\bf S}_A^3/\ZZ_{kq}$,
the $\gamma$-cycle fibration over ${\bf S}_A^2$.
The symmetry group $\SU(2)_B\times\U(1)_B$ acts on $L_{kq}$ as
isometry.
The interval $kq$ of $\wt P$ eigenvalues
in (\ref{kkmomcon}) is explained by
the $\ZZ_{kq}$ orbifolding by the operator (\ref{Zk}).
The fractional shift $q_{[a,b]}/q$
in (\ref{kkmomcon}) is interpreted as the contribution of
the Wilson line
\begin{equation}
\frac{q_{[a,b]}}{q}
=\frac{1}{2\pi}\oint_{{\bf x}_a^{\alpha\gamma}-{\bf x}_b^{\alpha\gamma}}C_3
\quad\mod 1,
\label{kktorsion}
\end{equation}
where $C_3$ is the three-form field in the $11$-dimensional supergravity.
For this relation to be acceptable,
the torsion must be quantized by
\begin{equation}
\frac{1}{2\pi}\oint_{{\bf x}_a^{\alpha\gamma}-{\bf x}_b^{\alpha\gamma}}C_3
\in\frac{1}{q}\ZZ.
\label{conskk}
\end{equation}
The geometry of the internal space, however, does not
guarantee (\ref{conskk}).
Because the $3$-cycle ${\bf x}_a^{\alpha\gamma}-{\bf x}_b^{\alpha\gamma}$
generates $\ZZ_{kq}$ subgroup of the homology $H_3(X_7,\ZZ)$,
the right hand side
in (\ref{kktorsion})
is quantized by
\begin{equation}
\frac{1}{2\pi}\oint_{{\bf x}_a^{\alpha\gamma}-{\bf x}_b^{\alpha\gamma}}C_3
\in\frac{1}{kq}\ZZ,
\label{conskk2}
\end{equation}
but this is not sufficient to guarantee (\ref{conskk}).

The quantization (\ref{conskk}) is explained in the following way.
The discrete torsion of $C_3$ represents the fractional M2-branes\cite{Aharony:2008gk,Imamura:2008ji}.
Because we consider the case in which all the gauge groups
are $\U(1)$ and there are no fractional M2-branes,
we should restrict the torsion to ones
corresponding to such situations.
In \cite{Imamura:2008ji} the relation between
the torsion and the numbers of fractional M2-branes
in the case of ${\cal N}=4$ Chern-Simons theories
is studied,
and the result shows that
the absence of the
fractional M2-branes requires
\begin{equation}
\frac{1}{2\pi}\int_{{\bf v}_a^{\alpha\gamma}}C_3\in\ZZ,\quad
\frac{1}{2\pi}\int_{{\bf w}_{\dot a}^{\alpha\gamma}}C_3\in\ZZ.
\label{nofrac}
\end{equation}
Because ${\bf v}_a-{\bf v}_b=-q({\bf x}_a-{\bf x}_b)$
follows from (\ref{vwdef}),
the first quantization condition in (\ref{nofrac}) guarantees
(\ref{conskk}).

We can easily see that
the spectrum of $\wt T_3$ in (\ref{wtt3mm}) is reproduced
by a scalar wave function of the M2-brane collective motion
in the Lens space $L_{kq}$.
The spherical harmonics in $L_{kq}$ is
obtained from ${\bf S}^3$ spherical harmonics
$Y_{l,m,m'}$
by restricting $\wt P$ eigenvalues by (\ref{kkmomcon}).
$Y_{l,m,m'}$ has three indices,
one angular momentum $l$ and two magnetic quantum numbers $m$ and $m'$,
which satisfy
\begin{equation}
-l\leq m,m'\leq l.
\label{mmbound}
\end{equation}
$Y_{l,m,m'}$ belongs to the spin $(l,l)$ representation
of the ${\bf S}^3$ rotational group $\SO(4)\sim\SU(2)^2$,
and $m$ and $m'$ are acted by two $\SU(2)$ factors separately.
Let us choose $\SU(2)_B\times\U(1)_B\subset\SO(4)$
so that $\SU(2)_B$ and $\U(1)_B$ act on $m$ and $m'$, respectively.
Then $m'$ is identified with the half of the $\U(1)_B$ charge $\wt P$,
and $(l,m)$ with the $\SU(2)_B$ quantum numbers.
The inequality (\ref{mmbound}) means that
for a given $\wt P$,
allowed $\SU(2)_B$ angular momenta are
\begin{equation}
l=
\frac{1}{2}|\wt P|,\quad
\frac{1}{2}|\wt P|+1,\quad
\frac{1}{2}|\wt P|+2,\quad
\ldots,
\end{equation}
and this correctly reproduce (\ref{wtt3mm}).

Because $\SU(2)_A\times\U(1)_A$ acts on the $L_{kq}$
as transverse rotations,
the corresponding charges $T_3$ and $P$ should be interpreted as
spins of M2-branes.
For $m=0$, we interpreted this above as the R-charge of a
scalar field on the D6-branes.
Thus, it seems natural to expect that the spectrum with $m\geq1$
is also reproduced as the spin of the M2-brane in excited states.

Because the charge $P$,
which is the D-particle charge from the type IIA perspective,
vanishes,
it may be possible to regard
the excited M2-brane as an excited open string on
the D6-branes.
Indeed, if we approximate the D6-branes by the flat ones,
there is the unique lowest energy state for each $T_3\geq1$,
and this seems consistent with (\ref{t3m}).
This is of course very rough argument
because the D6-branes and the background geometry
have large curvature.

\section{Conclusions and discussions}\label{conc.sec}
In this paper we computed the conformal dimensions
and the global $\U(1)$ charges of primary monopole operators $\MM_{ab}$
which carries non-diagonal magnetic charges corresponding to
roots of the $\SU(p)$ algebra.
In addition to the non-diagonal monopole charges,
the operators are labeled by
three integers $d$, $m\geq0$, and $n\geq0$.
We identified these operators with M2-branes wrapped on two-cycles in
the internal space, and
we interpreted $d$ and $n$ with the quantum numbers associated with
the orbital motions of wrapped M2-branes.
We also proposed that the quantum number $m$ may represent the
spin of excited M2-branes.

In this paper we considered Abelian Chern-Simons theories only.
It is important to generalize the analysis to
non-Abelian case.
Then, we can take the large $N$ limit,
and more reliable analysis on the gravity side
becomes possible.
Furthermore, such a generalization
enables us to
study the relation between general
discrete torsion and spectrum of monopole operators.
If we take a general discrete torsion quantized by (\ref{conskk2}),
the quantization of the momentum $\wt P$
is changed.
This should be realized as the monopole spectrum.

More challenging issue is the generalization to
theories with less supersymmetries.
In the case of ${\cal N}\leq 2$,
the large quantum corrections are expected
and the R-charges may be largely corrected.
On the gravity side, two-cycles have in general
non-vanishing area, and in such a case the computation
on the gravity side predicts the conformal dimension
of order $N^{1/2}$.
It would be very interesting if we could explain this behavior
as a result of dynamics in Chern-Simons theories.

\section*{Acknowledgements}
I would like to thank S.~Yokoyama for valuable discussions.
This work was supported in part by
Grant-in-Aid for Young Scientists (B) (\#19740122) from the Japan
Ministry of Education, Culture, Sports,
Science and Technology.

\end{document}